\documentstyle[floats,aps]{revtex}
\begin{document}
\input epsf

\title{Colliding black holes: how far can the close approximation go?}

\author{Reinaldo J. Gleiser$^1$, Carlos O. Nicasio$^{1,2}$,
Richard H. Price$^{3}$, Jorge Pullin$^4$}
\address{1. Facultad de Matem\'atica, Astronom\'{\i}a y F\'{\i}sica,
Universidad Nacional de C\'ordoba,\\ Ciudad
Universitaria, 5000 C\'ordoba, Argentina.}
\address{
2. Center for Gravitational Physics and Geometry, Department of
Physics,\\
The Pennsylvania State University, 
104 Davey Lab, University Park, PA 16802}
\address{
3. Department of Physics, University of Utah, Salt Lake City, Utah
84112.}

\maketitle
\begin{abstract}
We study the head-on collision of two equal-mass momentarily
stationary black holes, using black hole perturbation theory up to
second order.  Compared to first-order results, this significantly
improves agreement with numerically computed waveforms and
energy. Much more important, second-order results correctly indicate
the range of validity of perturbation theory. This use of
second-order, to provide ``error bars,'' makes perturbation theory a
viable tool for providing benchmarks for numerical relativity in more
generic collisions and, in some range of collision parameters, for
supplying waveform templates for gravitational wave detection.
\end{abstract}

\pacs{4.30+x}

\vspace{-9cm} 
\begin{flushright}
\baselineskip=15pt
CGPG-96/9-1  \\
gr-qc/9609022\\
\end{flushright}
\vspace{7cm}

The head-on collision of two momentarily stationary black holes has
become a classic problem in general relativity.  This is in part due
to the fact that it is simple, when compared to more generic
collisions, yet can be used as a testbed for ideas that will later be
applied in the astrophysically more relevant case of an inspiralling
coalescence.  The problem is far from academic.  The
near advent of gravitational wave detectors, like the LIGO or VIRGO
projects, that should be able to  measure gravitational radiation from
colliding black holes, is a strong motivation for obtaining
accurate waveform templates as soon as possible.  Because one expects
that most of the signal will be hidden by noise in a gravitational
wave detector, knowing the templates accurately can mean the
difference between detecting and not detecting \cite{Fi} gravitational
waves.

An important approach to the problem has been with numerical
relativity, pioneered by Smarr \cite{Sm} in the 70's.  The state of
the art in these techniques yields a fairly accurate computation of
radiated energies and waveforms for head-on (axisymmetric)
coalescences\cite{NCSA}, but further development of numerical
relativity will be needed to generalize computations to cases with
less symmetry.  Recently \cite{PrPu,2bhcollab}, perturbative
techniques applied to this problem were found to give remarkably good
results when the holes were initially close.  By treating the
collision of close black holes as the dynamics of a single distorted
black hole, one can use the well understood machinery of black hole
perturbation theory to treat the problem.  Better yet, generalization
of these techniques to collisions without axisymmetry entails only
minor modifications.

A key missing element in perturbative close-approximation calculations
is that one cannot tell {\it a priori} over what range it is
valid. (How close is ``close enough''?) The main point of this article
is to show that this missing element is provided by second-order
perturbation theory, which has recently been developed
\cite{GlNiPrPu}, largely for this purpose.  A higher-order calculation
is characterized by a number of interesting technical issues (such as
gauge choices of different order), and by considerable
complexity. Here we want only briefly to describe the nature of the
computation, and to argue the importance of the first results.  We
initially specialize to the details (e.g., even-parity only) of a
head-on collision of two non-spinning equal-mass holes starting with
Misner\cite{Mi} initial data. This is done for definiteness in the
description, and because this is the case which can most thoroughly be
compared with the numerical results.

The perturbation calculation starts with an expansion of the metric,
in terms of a perturbation parameter $\epsilon$, in the form
$g_{\mu\nu}=g_{\mu\nu}^{\rm Schw} + \epsilon
h_{\mu\nu}^{(1)}+\epsilon^2 h_{\mu\nu}^{(2)}+{\cal O}(\epsilon^2)$,
where $g_{\mu\nu}^{\rm Schw}$ is the usual Schwarzschild metric, for a
spacetime of mass $M$. Work on perturbation theory, more than twenty
years ago, showed that for even-parity perturbations all the physical
information in each $\ell$-pole of the first-order perturbations
$h_{\mu\nu}^{(1)}$ can be encoded into a single ``Zerilli function''
$\psi^{(1)}_\ell$ that is a combination of the $h_{\mu\nu}^{(1)}$ and
their derivatives\cite{zerilli,moncrief}. This function satisfies the
Zerilli equation, a simple, linear wave-like equation.  If we
construct essentially the same combination of $h_{\mu\nu}^{(2)}$ to
form a second-order Zerilli function $\psi^{(2)}_\ell$, the Einstein
vacuum equations, to second order in $\epsilon$, guarantee that
$\psi^{(2)}$ will satisfy a wave equation
\begin{equation}\label{zerwsource}
-\partial_t^2 \psi^{(2)}_\ell +\partial_{r^*}^2 \psi^{(2)}_\ell
+V_\ell(r)
\psi^{(2)}_\ell ={\cal S}_\ell\ ,
\end{equation}
which differs from the usual Zerilli equation only in the presence of
a nonzero ``source'' term on the right\cite{notnew}.  In this equation,
$r^*\equiv r+2 M \ln(r/2M-1)$ and $V_\ell (r)$ is given, e.g., in
\cite{zerilli,moncrief}.  The new feature for second-order, 
the source term ${\cal S}_\ell$, is a quadratic function of
$\psi^{(1)}_{\ell}$. In the case of the head-on collision, though not
more generally, it turns out that only the $\ell=2$ first-order terms
contribute to the quadrupole source ${\cal S}_2$. For this case the
source is explicitly given in
\cite{GlNiPrPu}.

There are two ways in which $\psi^{(2)}_{\ell}$ is not exactly the
same combination of the $h_{\mu\nu}^{(2)}$, as $\psi^{(1)}_{\ell}$ is
of the $h_{\mu\nu}^{(1)}$. First, for technical
reasons\cite{GlNiPrPu}, $\psi^{(2)}_{\ell}$ is actually equivalent to
the time derivative of the second-order Zerilli function. Second,
$\psi^{(2)}_{\ell}$ is not unique; one can redefine it up to terms
quadratic in the first-order perturbations.  The choice made in
\cite{GlNiPrPu} ensures that the source term is such that it yields a
well behaved $\psi^{(2)}_2$ in the radiation zone. (Here
$\psi^{(2)}_2$ corresponds to $\chi$ in the notation of
\cite{GlNiPrPu}.)

For the head-on collision problem, the perturbation expansion
parameter $\epsilon$ is a measure of the initial separation of the
holes. The particular choice we make is the parameter denoted
$\kappa_2$ in Ref.~\cite{PrPu} that occurs rather naturally in the
description of the problem.  It is straightforward to expand the
Misner initial data in this parameter and to find $h_{\mu\nu}^{(1)}$
and $h_{\mu\nu}^{(2)}$, from which we derive the initial conditions
for the dominant $\ell=2$ perturbations,
\begin{eqnarray}
\psi^{(1)}_2|_{t=0} &=& {128\over 3} {M^3 r\over (2 r+3 M)} {[7
\sqrt{r}+5 \sqrt{r-2 M}]\over [\sqrt{r}+\sqrt{r-2 M}]^5}\\
\partial_t \psi^{(1)}_2|_{t=0} &=&0\\
\psi^{(2)}_2|_{t=0} &=& -{131072\over 7} {M^6\over r (\sqrt{r}+\sqrt{r-2
M})^{10}}\label{initpsi2}\\
\partial_t \psi^{(2)}_2|_{t=0} &=&{16384 \over 7} {M^6 [5 (r-M)\sqrt{r-2M} +
19 \sqrt{r} (r-2 M)]\over (\sqrt{r}+\sqrt{r-2 M})^{10} (2 r +3
M)}\label{deriv}\ .
\end{eqnarray}
(The surprising  nonzero time derivative in (\ref{deriv}) 
arises partly because  $\psi^{(2)}_{2}$ is
actually the time derivative of the second-order Zerilli function, and partly
due to the quadratic terms that were added to  $\psi^{(2)}_{2}$ to make it
well behaved in the radiation zone \cite{tt}.)
The solution for $\psi^{(1)}_2$, and therefore the source term for
(\ref{zerwsource}), is known numerically from the solution to the
first-order problem\cite{PrPu}. This source and the initial data of
(\ref{initpsi2}) and (\ref{deriv}) complete the specification of the
problem defined by (\ref{zerwsource}).

With the solution of that problem for $\psi^{(1)}_2$, from a simple
finite differencing scheme (and with $\psi^{(1)}_1$ from the
first-order solution), the radiated power is found from
\begin{equation}\label{power}
{\rm Power}= {3\over 10} \left\{ \epsilon {\partial \psi^{(1)}_2 \over
\partial t} +
\epsilon^2 
\left[
 \psi^{(2)}_2  +
{1 \over 7} {\partial \over \partial t} \left( \psi^{(1)}_2
{\partial \psi^{(1)}_2 \over \partial
t}\right)\right]\right\}^2\ ,
\end{equation}
and the energy radiated is the time integral of this expression.  The
expression in brackets $\left\{\right\}$ is proportional to our second-order
radiation wavefunction, and omits yet-higher-order corrections.  In
applying (\ref{power}), it would be inconsistent to keep the terms of
order $\epsilon^4$, which are of higher order than omitted terms, so
we compute energy only to order $\epsilon^3$.

\begin{figure}[h]
\vskip 2cm
\epsfxsize=420pt \epsfbox{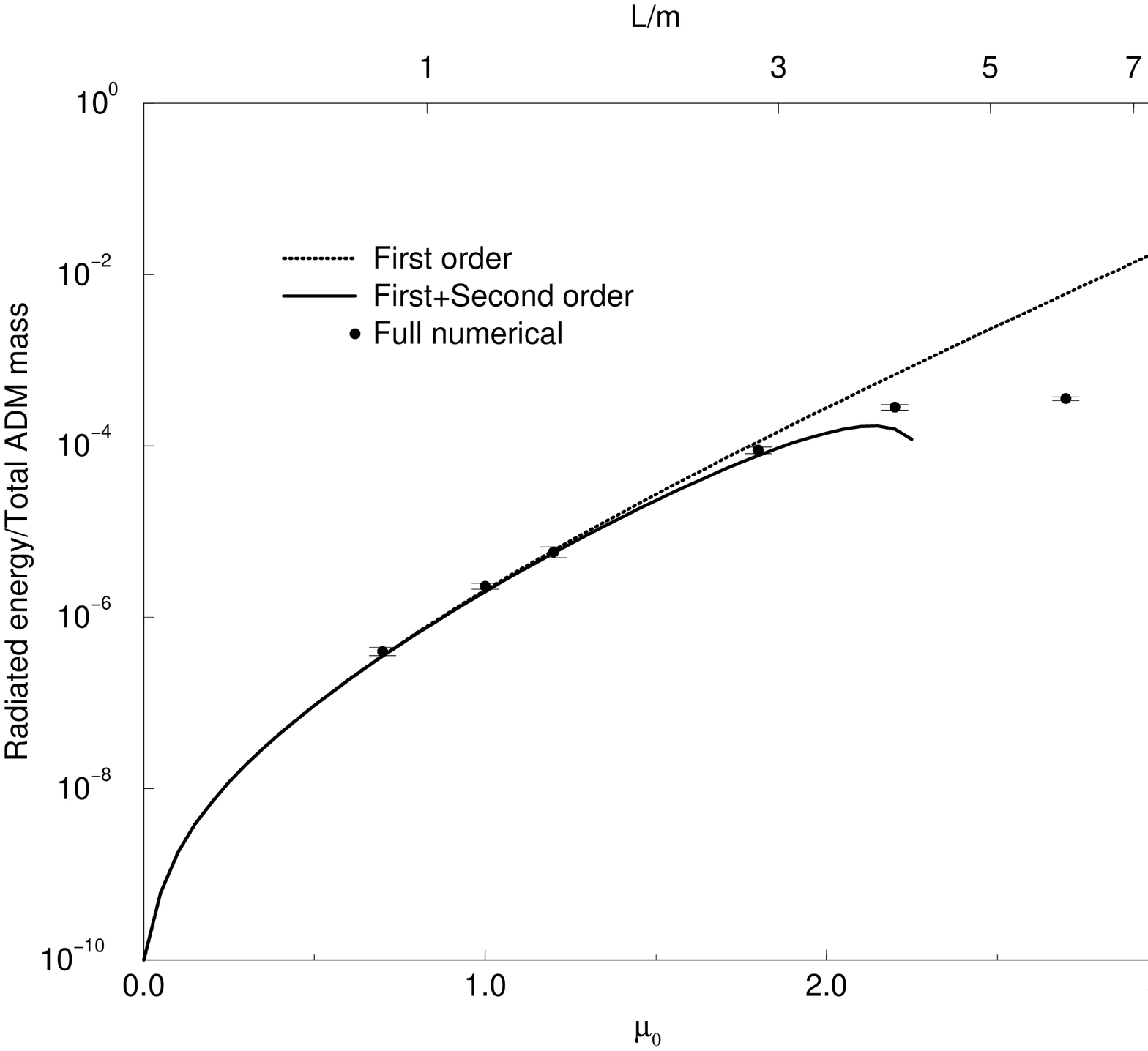}
\vskip 1cm
\caption{The total radiated energy in a black hole collision as a
fraction of the spacetime mass $M$ for different values of the initial
separation. The upper scale is in units of $L/m$, where $m$ is the
mass of each individual hole (inferred from the minimal area of the
throat for each hole); the bottom scale is in the parameter $\mu_0$
that appears in the Misner data.  The dotted curve is the first-order
result and the solid curve the result up to second order. For
$\mu_0\approx 2.2$ the second-order corrections are as large as the
first order result. The numerical results (with error bars) correspond
to the NCSA simulations of Ref.\ [3].}
\label{cac}
\end{figure}

Figure 1 shows the energy results (to order $\epsilon^3$) as a
function of the Misner\cite{Mi} parameter $\mu_0$. This figure
compares these second-order results with first-order results and with
those of numerical relativity, and illustrates the point we wish to
emphasize. If we had only the first-order calculations, we would know
that the predictions at very large separation parameter $\mu_0$ were
unreasonable, but we would not know how small $\mu_0$ must be for the
results to be reasonably accurate. By comparing the first-order
prediction with that of the second-order computation, we can infer
that perturbation theory is applicable up to $\mu_0$ of order 1.8 or
so. (The actual cutoff would depend on what level of accuracy one
demands.) In the case of the head-on collision from Misner data, this
prediction is subject to verification by comparison with the results
of numerical relativity. Figure 1 clearly shows that the prediction is
correct; perturbation theory is valid (i.e., agrees with numerical
relativity results) up to around $\mu_0=1.8$.  Second-order
calculations {\em do} in this case provide ``error bars'' for
perturbation calculations.  There is nothing specific about the
head-on problem that favors this outcome, and it is reasonable to
infer that that second-order results will provide the same ``error
bars'' in generic coalescences.

A mild caveat hangs over the application of this method to other
cases.  There is no unique division of perturbations into
orders\cite{AbPr}.  If, for example, one knew the exact radiated
energy for the head-on collision, for all values of $\mu_0$, one could
define a new expansion parameter $\epsilon$ that depends on $\mu_0$ in
such a way that first-order perturbation theory would be exact! More
worrisome, in any perturbation problem one could define an $\epsilon$
for which the first-order energy is not exact, but for which the
second-order perturbations are zero.  The ``error bars'' provided by
these second-order perturbations would be wildly
misleading\cite{however}. Such cases keep our method of error-bar
determination from being rigorous, but are a problem more in principle
than in practice. Anomalous behavior of an order of perturbations
arises only in atypical circumstances, usually when one contrives to
arrive at just such an atypical result. For a ``generic'' perturbation
parameter one expects the second-order error-bar method to be just as
useful as in Fig.\ 1.

\begin{figure}[h]
\epsfxsize=420pt \epsfbox{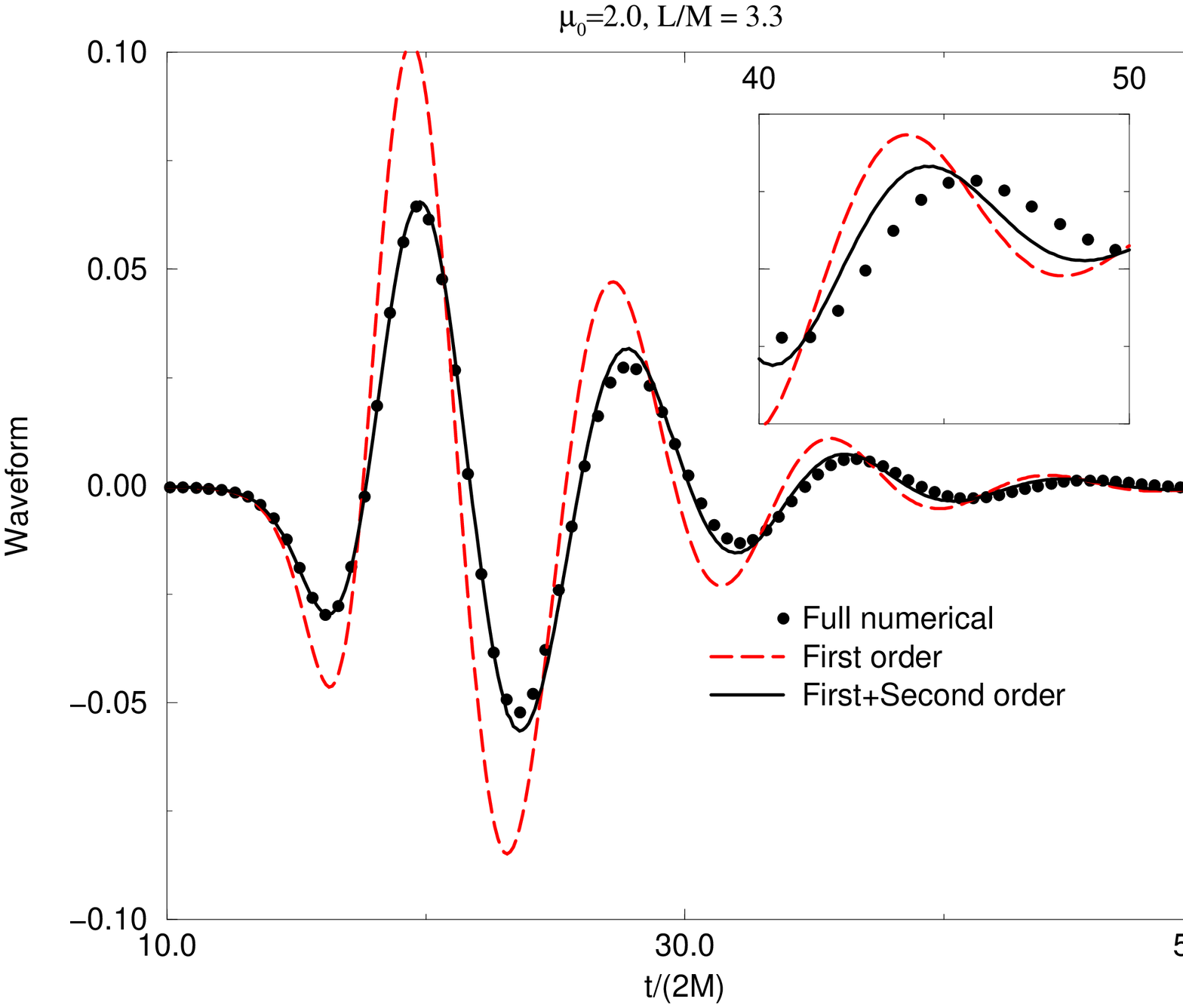}
\caption{ Waveforms for $\mu_0=2$.  Shown are the waveform for time
derivative of the first-order Zerilli function and that waveform with
the second-order correction added. This is compared with the
equivalent waveform from numerical relativity. We see a significant
improvement due to the use of second-order perturbation theory, even
for this case where the separation of the two black holes is
$L/M\approx 3.3$. The inset shows late oscillations. The period of the
numerical relativity result for these oscillations becomes larger than
the quasinormal period and is thought to be a numerical artifact
[5]. The second-order correct curve then is probably more accurate in
this region. The full numerical data sets have higher resolution than
the one depicted, we have omitted some data points for visual
clarity.}
\end{figure}

A secondary advantage of second-order computation is that it can give
improved accuracy of results. The difference between first-order
results and second-order-correct results is, of course, significant only for
cases in which perturbation theory is marginally applicable. But it is
just these cases for which the radiation is strongest and which will
typically be of greatest interest for comparison with numerical
relativity, or for considerations of detector design.  In Fig.\ 2 we
show a comparison of waveforms for the marginal case of $\mu_0=2$,
corresponding to an initial proper distance $L=3.3M$ between the
throats. At this separation there is no single all-encompassing
horizon surrounding both throats, so the  assumptions of the
close limit approximation are  violated. As expected, the first-order
result shows significant disagreement with the numerical relativity
result, but the result of our second-order computation is a waveform
in remarkably good agreement with the numerical relativity result.

The success is not only remarkable, it is  at first somewhat puzzling. The
waveform for second-order theory has an amplitude that is at most
around 10\% larger than the numerical relativity waveform, implying
that the second-order result for radiated energy should be around 20\%
larger than that from numerical relativity. In Fig.\ 1, however, the
second-order energy is about half the result of numerical relativity
at $\mu_0=2$. The explanation of this brings out an interesting point
about the computations. If we take the second-order waveform of Fig.\
2, square it, multiply it by the correct factor and integrate, we
calculate an energy which is, as it must be, about 20\% larger than
that from numerical relativity (i.e., the result of the same
operations on the numerical relativity waveform of Fig.\ 2). But this
procedure for computing second-order energy includes the formally
disallowed $\epsilon^4$ terms in (\ref{power}). The success of this
formally inconsistent procedure suggests that the omitted terms in
(\ref{power}) are small, so that the effect of omitting them is
unimportant. We must keep in mind that for a marginal case like
$\mu_0=2$ the expansion parameter is of order unity, so that ``order
in $\epsilon$'' is not necessarily the critical factor in the
importance of a term.  The omitted terms in (\ref{power}) turn out to
be the result of a nonlinear mixing of $\ell=2$ and $\ell=4$
contributions. The $\ell=4$ contribution for the problem is small,
quite aside from the fact that it is higher order in $\epsilon$. [See
Fig.\ 12 of Ref.~\cite{2bhcollab}; at $\mu_0=2$, the $\ell=4$ energy
is three orders of magnitude smaller than the $\ell=2$ energy.] Our
higher-order waveforms are more accurate than are implied by a
comparison of first-order and second-order-correct energies, but this
fact may be peculiar to the specific case of initial Misner data, and
not applicable to more general black hole collisions.

It is the study of more general collisions to which this work is
directed.  We have shown that we have a practical and reliable (though
not rigorous) index of the accuracy of perturbation theory, and hence
the basis for a consistent investigation of collisions that are of
interest as sources of detectable gravitational waves. With estimates
of accuracy established by higher-order calculations, perturbation
theory will give waveforms and energies that will check numerical
relativity codes; that can help in exploring the parameter space of
collisions for those cases of interest for greater numerical work;
that can in some cases provide guidelines for development of details of
detection schemes. Work is already underway on the application of this
technique to the head-on collision of initially moving and spinning
holes, and has started on the modifications that are needed when the
final hole is rapidly rotating.
\vskip 0.5cm

We wish to thank Peter Anninos for help in providing the full
numerical results from the NCSA group.  This work was supported in
part by grants NSF-INT-9512894, NSF-PHY-9423950, NSF-PHY-9507719, by
funds of the University of C\'ordoba, the University of Utah, the
Pennsylvania State University and its Office for Minority Faculty
Development, and the Eberly Family Research Fund at Penn State. We
also acknowledge support of CONICET and CONICOR (Argentina). JP also
acknowledges support from the Alfred P. Sloan Foundation through an
Alfred P. Sloan fellowship.

%%%%%%%%%%%%%%%%%%%%%%%
%%%%%%%%%%%%%%%%%%%%%%%
%%%%%%%%%%%%%%%%%%%%%%%


\begin{references}
\bibitem{tt} The transverse-traceless metric perturbation in the 
radiation zone is given by the expression within the curly braces in
equation (6).
\bibitem{however}Changing orders of perturbation in this way is guaranteed
only for a single function of the perturbation parameter (in our case,
the energy as a function of separation). The different sorts of
information contained in a waveform (phases, relative amplitudes of
different peaks, etc.) may not all be subject to simultaneous
manipulation in this way. Comparing first-order wave forms with
second-order-correct waveforms, therefore, may give a more certain
determination of the valid range of perturbation theory, especially in
cases where the waveforms have richer detail than for head-on
collisions.
\end{references}
\end{document}